\documentclass[%
amsmath,amssymb,
 aps,
pra,
 longbibliography,
 lengthcheck,%
]{revtex4-1}

\usepackage{amssymb}
\usepackage{amsmath}
\usepackage{graphicx}
\usepackage{epsfig}
\usepackage{mathtools}
\usepackage{dsfont}
\def\bea{\begin{eqnarray}}
\def\eea{\end{eqnarray}}

\def\ba{\begin{equation}\begin{array}{c}}
\def\ea{\end{array}\end{equation}}
\def\be{\ba\displaystyle}
\def\ee{\ea}

\begin{document}


\title{Quantum Boltzmann equation for a mobile impurity\\ in a degenerate Tonks-Girardeau gas}

\author{O. Gamayun$^{1,2}$}
\affiliation{$^1$ Lancaster University, Physics Department, Lancaster LA1 4YB, UK}
\affiliation{$^2$Bogolyubov Institute for Theoretical Physics, 14-b Metrolohichna str., Kyiv 03680, Ukraine}

\begin{abstract}

We investigate the large-time asymptotical behavior of a mobile impurity immersed in a degenerate Tonks-Girardeau gas.
We derive a correct weak-coupling kinetic equation valid for arbitrary ratio of masses of gas and impurity  particles.
When gas particles are either lighter or heavier than the impurity we find that our theory is equivalent to the Boltzmann theory with the collision integral calculated via the Fermi Golden Rule. On the contrary, in the equal-mass case, Fermi Golden Rule treatment gives false results due to not accounting for multiple coherent scattering events. The latter are treated by the ressummation of ladder diagrams, which leads to a new kinetic equation.
The asymptotic momentum of the impurity produced from this equation coincides with the result obtained by means of the Bethe ansatz.
\end{abstract}
\maketitle


\date{\today}

\section{\label{sec intro} Introduction}

The propagation of impurities in quantum liquids and gases has always attracted a lot of attention from researchers \cite{Khalat,Meyer58,prokof1993diffusion}.
Recently this interest has revived, with a focus on one-dimensional (1D) systems. This is partially due to the tremendous experimental progress in fabricating and manipulating ultracold atomic gases \cite{bloch2008many}, which has allowed the creation and manipulation of a single-impurity state in a 1D host gas of bosons \cite{sherson2010single,weitenberg2011single} and examination of its non-equilibrium dynamics \cite{palzer2009quantum,spethmann2012dynamics,catani2012quantum,fukuhara2013quantum,fukuhara2013microscopic}.

The theoretical interest in impurity propagation in 1D systems is due to the rich variety of unusual properties of the host liquids \cite{Giamarchi2003,imambekov2012one}. One of the prominent features is the substantial modification of the superfluidity (understood as the absence of friction force) in 1D liquids \cite{CauxReview}. Another intriguing
phenomenon is quasi-Bloch oscillations, comprised of impurity momentum oscillations in the presence of external force and, in this way, resembling the Bloch oscillations in an ideal crystal \cite{Gangardt2009,Gangardt2012,schecter2012dynamics}. The most puzzling phenomenon  concerning impurity motion, though, is
quantum flutter phenomenon discovered in Ref. \cite{mathy2012quantum} and further explored in Ref. \cite{knap2013quantum}.  The authors considered supersonic impurity injected into a 1D gas of hardcore bosons, also known as a Tonks-Girardeau gas. This gas is equivalent to the free-fermion system \cite{Girardeau1960}. At zero temperature host fermions form a Fermi sea in which impurity propagation is considered (a more general host was considered in \cite{knap2013quantum}). Numerical analysis in \cite{mathy2012quantum,knap2013quantum} clearly shows that the large-time asymptotic of the average momentum exhibits oscillations around some non-zero value. These results suggest the existence of an asymptotical steady state with a non-vanishing momentum of the impurity. The nature of this state was analyzed in \cite{BChGL2013,Burovski2014} where a complete analytical theory for the description of this state has been developed. The dependence of the asymptotic momentum $p_{\infty}$ on the initial momentum $p_0$ has been calculated in weak-coupling regime.
From the kinetic arguments it is clear that if the initial momentum is less than some critical $|p_0|<q_0$ then even a single act of scattering is prohibited by classical conservation laws and the asymptotic momentum of the impurity coincides with initial, $p_{\infty}=p_0$. This could be considered as a Landau-like criteria (see also Ref. \cite{Lychkovskiy2013} for taking into account interaction and getting non-perturbative bounds on $p_{\infty}$). If the initial momentum is higher than $q_0$, then after several scattering events, the impurity momentum drops below $q_0$ and further scattering stops. To calculate this asymptotical value semiclassical Boltzmann theory was invoked in Ref. \cite{BChGL2013}. The Boltzmann theory treatment heavily relies on the assumption that all the dynamics reduces to a sequence of pairwise collisions. This is applicable when the impurity mass is not equal to the host particle mass (non-equal masses case).
On the other hand, for equal masses, the impurity momentum drops below $q_0$ after the first scattering
forming a hole in the Fermi sea. And, unlike  the nonequal-mass case, the velocity of the hole equals to the velocity of the particle and they may experience
multiple coherent scattering processes leading to a resonant interaction between the impurity and the host.
When interaction between the impurity and the host particle is point-like, the equal masses case is integrable by means of the coordinate Bethe ansatz \cite{mcguire1965interacting}. This technique was used in \cite{BChGL2013} and it was obtained that
in the vanishing coupling constant limit, the asymptotic momentum
acquires the following nontrivial value:
\be\label{asymptotic momentum equal masses}
p_\infty=p_0-\theta\left(|p_0|-k_{\rm F}\right)\frac{p_0^2-k_{\rm F}^2}{2k_{\rm F}} \ln \frac{p_0+k_{\rm F}}{p_0-k_{\rm F}}\,.
\ee
Here $k_F$ is Fermi momentum, which in this case coincides with $q_0$.

Note that the large-time limit and the small coupling constant limit do not commute. Obviously, when the coupling constant is taken to zero first, then the dynamics is trivial and the asymptotical momentum is equal to the initial one, $p_\infty=p_0$. However, the result in Eq. \eqref{asymptotic momentum equal masses} is obtained when the large-time asymptotic of the momentum is calculated at a finite value of the coupling constant; subsequently, this value is taken to zero.

Note that the asymptotic momentum calculated within the Boltzmann theory approach is equal to
\begin{equation}
 p_{\infty}^{B} = p_0 - 2k_F\,\theta\left(|p_0|-k_F\right)  \left( \ln \frac{p_0+k_F}{p_0-k_F} \right)^{-1}\,.
 \label{Bolt}
\end{equation}
This clearly shows that, even at a vanishingly small coupling constant accounting for multiple scattering processes provides finite nonperturbative corrections to the final result.

In this manuscript we analyze this phenomenon in a systematic way. We derive a kinetic equation that describes the impurity momentum probability distribution from the Schwinger-Dyson equation in the Keldysh formalism. To solve this equation and find the asymptotic distribution we have to specify diagrams used in the impurity self-energy $\Sigma$ (collision integrals). Standard Boltzmann theory corresponds to the $\Sigma$ in the form of the single-bubble diagram, which is the lowest order expansion in the coupling constant. This approximation is equivalent to the Boltzmann equation with probability transitions computed by the Fermi Golden Rule. This approach works well in the nonequal-mass case, meaning that taking into account higher orders in $\Sigma$ results in higher order corrections in the final answer for the asymptotic probability distribution. At equal masses this is no longer true. Namely, the asymptotic distribution calculated from the bubble diagram at equal masses for the initial impurity momentum $p_0>k_F$ is found to be:
\be\label{1}
\left[n_p^B\right]^{\infty} = \frac{1}{Z^B_{p_0}} \frac{\theta(k_F-|p|)}{p_0-p}\,,
\ee
here $Z^B_{p_0}$ is an appropriate normalization constant. This result can be used to reproduce answer \eqref{Bolt}. However, if one takes into account, ladder diagrams and performs effective ressummation in $\Sigma$, one will find that the result for the asymptotic distribution is drastically changed even in leading order, providing
\be\label{2}
n_p^{\infty} = \frac{1}{Z_{p_0}} \frac{\theta(k_F-|p|)}{(p_0-p)^2}\,,
\ee
which leads to the correct result, \eqref{asymptotic momentum equal masses}, obtained by the means of the Bethe ansatz solution. If the mass ratio is far enough from unity, we find that
ladder effects are suppressed and Boltzmann theory is applicable. Also, we are able to write an effective Boltzmann-like equation in the equal-mass case
replacing transitions rates obtained by the Fermi Golden Rule with those calculated from ladder diagrams. Using this equation we consider the dynamics of
the external force applied to the impurity. Such considerations are usually extremely difficult in integrable systems and few analytical results have been obtained so far. It is straightforward to include the finite temperature and trap potential in our approach, but we postpone this to separate consideration.

The plan of the paper is as follow: in the next section we describe the physical system and introduce the main notations.
In Sec. \ref{Dyson} we derive a general kinetic equation to describe the impurity and present the results for the kinetic equation with
specific ladder contributions to the impurity self-energy.  In Sec. \ref{Solution} we determine the expression for the Green function,
analyze the asymptotic distributions and obtain the main results of our paper, in particular, Eq. \eqref{2}.
Section \eqref{Force} is devoted to the consideration of an equal-mass system with applied constant force. Finally, a short summary and discussion are present in Sec. \eqref{Discussion}.

\section{Physical system and general Setup}

As mentioned in Sec. \eqref{sec intro}, we consider an impurity particle immersed in a Tonks-Girardeau gas. This gas is equivalent to free fermions \cite{Girardeau1960} and we will use this fermionic description of the host. The Hamiltonian of our system in the second quantization language reads
\be\label{Ham}
H =\sum_p \epsilon_p a^+_pa_p + \sum\limits_p E_pb^+_pb_p + g \sum\limits_{p,q,s} b^+_{p-q}b_pa^+_{s+q}a_s
\ee
Here the operators $a_p$ corresponds to the host fermions and $b_p$ to the impurity. The Hilbert space for the impurity is reduced to be one particle, i.e. it consists of the vacuum state $|0\rangle$ and linear combinations of the one particle states $b_{p}^+|0\rangle$ only. In this case it is not necessary to specify the statistics of the operators $b_p$. But, for certainty, whenever needed, we assume that the impurity is a fermion, presuming that all effects of statistics will cancel out in the final answers. The initial state of the whole system $|{\rm in}\rangle $, for the sake of simplicity, is taken to be a product state of the impurity at a given momentum $p_0$ and host particles in the Fermi sea state that is defined by momentum $k_F$: $|{\rm in}\rangle = b^+_{p_0} |0\rangle |FS\rangle $.
The spectrum of particles is assumed to be $\epsilon_p = p^2/2m_h$ and $E_p=p^2/2m_i$, though some results presented below are valid for an arbitrary spectrum. The system is assumed to have periodic boundary conditions with period $L$, and we denote  $\sum_p=L/2\pi \int dp$, and set  $\hbar=1$. The strength of the interaction is characterized by the dimensionless coupling constant $\gamma\equiv (gL/2\pi) m_h/k_F$ and all final answers depend on $\gamma$ only. Nevertheless, we keep $g$ for convenience in intermediate calculations. It is natural to
measure all momenta in units of $k_F$ and all masses in the mass of the host particle $m_h$. From now on we set $k_F=1$ and $m_h=1$. So the impurity mass is now given by the ratio $\eta = m_i/m_h$ and $\gamma=gL/2\pi$.

The long-time evolution from the the initial state $|{\rm in}\rangle $ is our primary concern. In particular, we would like to know the impurity momentum distribution probability, defined as
\be\label{prob}
n_p(t) = \langle {\rm in}|e^{itH}b_p^+b_pe^{-itH}|{\rm in} \rangle.
\ee
The initial distribution is given by the formula
\be
n_p(0)= \delta_{p,p_0}\,.
\ee
The most natural approach for evaluating \eqref{prob} is to use the Keldysh formalism technique \cite{Keldysh}.
Following standard procedures \cite{Kamenev,Altland}, we introduce the operators $\alpha_p$ and $\beta_p$ which are identical to $a_p$ and $b_p$, but have specific time ordering (live on a different contour). "Greek" operators precede "Latin" operators at any values of time, which runs from 0 to $\infty$ and are  anti-time-ordered among themselves, while "Latin" operators are time-ordered.
We perform Keldysh rotation by means of the matrices
\be
U = \frac{1}{\sqrt{2}}\left(
                           \begin{array}{cc}
                             1 & 1 \\
                             1 & -1 \\
                           \end{array}
                         \right),\,\,\tilde{U} =\frac{1}{\sqrt{2}}\left(
                           \begin{array}{cc}
                             1 & -1 \\
                             1 & 1 \\
                           \end{array}
                         \right),
\ee
introducing the new variables:
\be
\chi_k=U \left(
                                  \begin{array}{c}
                                    b_k \\
                                    \beta_k \\
                                  \end{array}
                                \right),\,\,\,\,\,\,\,
                                \bar{\chi}_k = \tilde{U}\left(
                                  \begin{array}{c}
                                    b_k^+ \\
                                    \beta_k^+ \\
                                  \end{array}
                                \right),
                                \ee
                                \be
                                \psi_k = U\left(
                                  \begin{array}{c}
                                    a_k \\
                                    \alpha_k \\
                                  \end{array}
                                \right),\,\,\,\,\,\,\,
                                \bar{\psi}_k =\tilde{U}\left(
                                  \begin{array}{c}
                                    a_k^+ \\
                                    \alpha_k^+ \\
                                  \end{array}
                                \right).
\ee
The interaction term takes the form:
\be\label{int}
H_{\rm int} = \frac{g}{2} \sum\limits_{a=0,1} \sum\limits_{k,q,s} \bar{\chi}_{k-q}\sigma_{1-a}\chi_k\bar{\psi}_{s+q}\sigma_{a}\psi_s,
\ee
with
\be
\sigma_0 = \left(\begin{array}{cc}
                             1 & 0 \\
                             0 & 1 \\
                           \end{array}
                         \right),\,\,\,\,\,\,\,
                         \sigma_1 = \left(\begin{array}{cc}
                             0 & 1 \\
                             1 & 0 \\
                           \end{array}
                         \right)
\ee
The bare impurity Green function takes the form
\be
\langle {\rm in}| T\chi_k(t_1)\otimes\bar{\chi}_p(t_2)  |{\rm in}\rangle \equiv \delta_{kp} \mathds{G}^{(0)}_pe^{-i(t_1-t_2)E_p},
\ee
where
\be\label{Green1}
 \mathds{G}^{(0)}_p = \left(
\begin{array}{cc}
 G_p^+ & G_p^K \\
0 & -G_p^-
\end{array}
\right) = \left(
\begin{array}{cc}
 \theta(t_1-t_2) & 1-2\delta_{p,p_0} \\
0 & -\theta(t_2-t_1)
\end{array}
\right)
\ee
is the quantity that we henceforth refer to as the Green function. Our definition differs from the usual one by prefactor $e^{-i(t_1-t_2)E_p}$, which in momentum space, corresponds to shifting the energy shell to zero. The analogous Green function for host fermions reads
\be\label{GreenF}
\mathds{F}_p =\left(
\begin{array}{cc}
 \theta(t_1-t_2) & {\rm sgn} (|p|-k_F) \\
0 & -\theta(t_2-t_1)
\end{array}
\right)
\ee
We see that the initial conditions enters through the Keldysh part of the Green function (the upper-right-corner element).
Therefore, it is useful to combine the diagonal elements, introducing the Feynman Green function:
\be\label{Feynman}
G_p(t_1,t_2) = \theta(t_1-t_2)G_p^+(t_1,t_2) + \theta(t_2-t_1)G_p^-(t_1,t_2).
\ee
Further, it is also useful to consider the following combination instead of $G^K_p$:
\be\label{prob1}
W_p(t_1,t_2) = \frac{G_p(t_1,t_2)-G^K_p(t_1,t_2)}{2}\,.
\ee
The probability distribution is given by the following formula, \eqref{prob}, as:
\be
n_p(t) = W_p(t,t);
\ee
therefore, we refer to quantity \eqref{prob1} as the generalized probability distribution.
It is of order $1/L$, contrary to the Feynman Green Function \eqref{Feynman},
which is of order $1$:
\be
\frac{1}{L}\sim W_k \ll G_k \sim 1\,.
\ee
One can show that because of this 'separation of scales', the Feynman Green function remains translational invariant in limit $L\to\infty$ even though it is not a vacuum average. This is easily understood because the one-particle contribution of the impurity produces only $1/L$ effect compared to the vacuum state:
\be
G_k(t_1,t_2)=  G_k(t_1-t_2) + O(1/L)
\ee
This can be considered as a low-density approximation, which is absolutely applicable here since we are dealing with a
single impurity and a thermodynamically large amount of host particles. The generalized probability distribution, however, retains its essential dependence on both times. This distribution satisfies the quantum Boltzmann equation which we derive in the next section.

\section{\label{Dyson}Quantum Boltzmann Equation}
\begin{figure}
\includegraphics[scale=0.35]{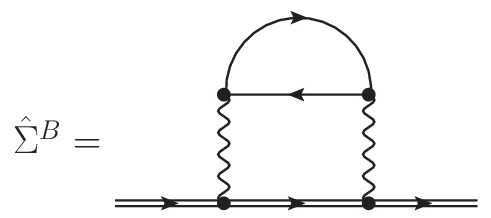}
\caption{
\label{LadderB}
Diagram of self-energy that corresponds to the Boltzmann approximation}
\end{figure}
Let us denote the self-energy of a particle, which is given by all appropriate one-particle irreducible diagrams, as $\hat{\Sigma}$. The full Green function,  \eqref{Green1}, is given by
\be\label{Dyson1}
[\mathds{G}]^{-1}= [\mathds{G}^{(0)}]^{-1}- \hat{\Sigma}
\ee
The self-energy maintains the same matrix structure as the Green function \cite{Kamenev}:
\be
\hat{\Sigma}_p(t_1,t_2) =
                                                            \left(
                                                              \begin{array}{cc}
                                                                \Sigma_p^+(t_1,t_2) & \Sigma_p^K(t_1,t_2) \\
                                                                0 & -\Sigma_p^-(t_1,t_2)\\
                                                              \end{array}
                                                            \right)
\ee
In the same manner as for the Green function we may introduce the following notations:
\be\label{22}
\Sigma_p(t_1,t_2) = \theta(t_1-t_2)\Sigma_p^+(t_1,t_2)  +
 \theta(t_2-t_1)\Sigma_p^-(t_1,t_2)
\ee
\be\label{23}
\sigma_p(t_1,t_2)  =\frac{\Sigma_p^K(t_1,t_2)-\Sigma_p(t_1,t_2)}{2}
\ee
\begin{figure}[t]
\includegraphics[scale=0.5]{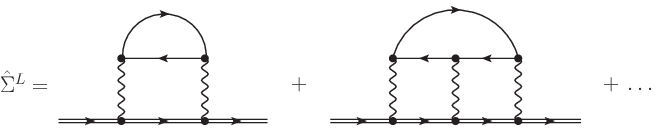}
\caption{
\label{LadderF}
The ladder diagram for self-energy that takes into account multiple impurity hole scattering events}
\end{figure}
Analogously as for the Green functions we have $\Sigma_p\sim 1$ and $\sigma_p\sim 1/L$. In the $L\to\infty$ limit $\Sigma_p$ is translational invariant, $\Sigma_p(t_1,t_2) = \Sigma_p(t_1-t_2)$ so Eq. \eqref{Dyson1} transforms into a system on two
integral equations,
\be\label{G}
G_p(\tau) = 1 + \int\limits_{0}^{\tau}dt\int\limits_{t}^{\tau} dt_1 \Sigma_p(t_1-t)G_p(t)
\ee
\begin{multline}\label{W}
W_p(\tau_1,\tau_2) = \omega^{(0)}_pG_p(\tau_1)G^*_p(\tau_2)+ \\
+ \int\limits_{0}^{\tau_1}dt_1 \int\limits_0^{\tau_2}dt_2  G_p(\tau_1-t_1)\sigma_p(t_1,t_2)G^*_p(\tau_2-t_2);
\end{multline}
here $n^{(0)}_p = \delta_{p,p_0}$, and a superscript asterisk indicates complex conjugation. For derivation of these equations the fact that $G_p(-\tau)=G^*_p(\tau)$ was used.
We note that Eqs. \eqref{G} and \eqref{W} can be self-consistently considered for positive times only. Therefore we may apply the Laplace transformation to these equation and obtain
\be\label{GL}
G_p(\lambda)^{-1} = \lambda -\Sigma_p(\lambda)
\ee
\be\label{QBE}
W_{p}(\lambda_1,\lambda_2) = G_p(\lambda_1)\left(n_p^{(0)}+\sigma_p(\lambda_1,\lambda_2)\right)G^*_p(\lambda_2)
\ee
One easily notes that, in leading $1/L$ order, $\Sigma_p$ does not depends on $W_p$, while $\sigma_p$ is dependent in a linear way. Therefore, Eqs. \eqref{G} and \eqref{GL} are
self-consistent and actually describe the vacuum Green function. Once this function is found, then Eqs. \eqref{W} and \eqref{QBE} present the linear integral equation on generalized probability distribution, which we call the Quantum Boltzmann Equation (QBE) in time and $\lambda$ space, respectively. The fact that the QBE is linear is the essence of the 'low-density approximation', which as discussed above, is exact in the thermodynamical limit ($L\to\infty$).

Equations \eqref{GL},\eqref{QBE} are valid for any systems in any dimensions. But to make them comprehensive we have to specify the self-energy pertinent for our system.
To do this we use the diagrammatic approach, which seems the most suitable in the perturbative limit.
Notations in all diagrams are as follows:
a wiggly line corresponds to interaction \eqref{int}, a straight line corresponds to host propagators \eqref{GreenF} and
a double-line corresponds to the impurity propagators, \eqref{Green1}. Because of the one-particle Hilbert space each diagram can contain no more than one
double-line.
In the leading order in the coupling constant $\hat{\Sigma}$ is given by the diagram in Fig. \ref{LadderB}.

This diagram describes an act of single scattering of the impurity on the host particle and leads to the Boltzmann theory based on the semiclassical Fermi Golden Rule. However, for equal masses even for
purely kinematic reasons one might expect that further scattering of the impurity on the hole is significant.
Indeed, when the velocities of the impurity and hole coincide, the resonant amplification of the interaction, heuristically, can be seen from the coordinate form of the interaction in Hamiltonian \eqref{Ham}: $V^{\rm int} \sim \gamma \delta(X^{\rm impurity}-X^{\rm hole}) \sim  \gamma \delta(t(v^{\rm impurity}-v^{\rm hole}))\sim \gamma \delta(0)\gg \gamma$. To take into account such effects we also consider the so-called ladder diagrams presented in Fig. \ref{LadderF} for arbitrary $\eta$. These diagrams describe interaction of the impurity with hole when they move along under the Fermi Sea. We find that each ladder diagram is proportional to $(\gamma/|\eta-1|)^n$ where $n$ is number of internal wiggly lines. All other diagrams apart from possible powers of $\gamma/|\eta-1|$ contains also higher orders in $\gamma$. Therefore, to obtain meaningful descriptions of the physical processes at $\eta\to 1$ ladder diagrams should be resummed.  This situation is similar to the resummation in  quantum field theories with a large number of fermion flavors, the so-called $1/N$ expansion, where the leading in $N$ expression is obtained after summing up all diagrams with planar topology \cite{Coleman}. So ladder diagrams provide the leading expansion in coupling constant $\gamma$ and all orders expansion in parameter $\gamma/|\eta-1|$. Below we clarify this statement analytically.

We calculate $\hat{\Sigma}_p$ from the corresponding diagrams and then determine $\Sigma_p$ and $\sigma_p$ using definitions \eqref{22} and  \eqref{23}. For the ladder diagrams (Fig. \ref{LadderF}) the result reads
\be\label{Glambda}
\Sigma^L_p(\lambda) = (ig)^2 \sum\limits_{|k|>1} \frac{\sum\limits_{|q|<1} G_{p-k+q}(\lambda+iE_{kq})}{1-ig\sum\limits_{|q|<1} G_{p-k+q}(\lambda+iE_{kq})}
\ee
\begin{widetext}
\begin{equation}\label{Slambda}
\sigma^L_p(\lambda_1,\lambda_2) = g^2\sum\limits_{|s|<1,|k|>1} \frac{W_{p-s+k}(\lambda_1+iE_{sk},\lambda_2-iE_{sk})}{\left(1-ig \sum\limits_{|q|<1} G_{p-s+q}(\lambda_1+iE_{sq})\right)\left(1+ig \sum\limits_{|q|<1} G^*_{p-s+q}(\lambda_2-iE_{sq})\right)}
\end{equation}
\end{widetext}
where
$E_{q_iq}$ is the transferred energy:
\be
E_{q_iq} = E_{p-q_i+q}-E_p+\epsilon_{q_i}-\epsilon_q\,
\ee
which for quadratic dispersions is equal to:
\be
E_{q_iq} = \frac{q_i-q}{\eta}\left(q_i\frac{\eta+1}{2}+q\frac{\eta-1}{2} - p \right)\,.
\ee
It retains the dependence on $p$ which is the incoming impurity momentum.

The Boltzmann diagrams (Fig. \ref{LadderB}) corresponds to the lowest $g$ orders that come from the ladder diagrams, namely,
\be\label{GlambdaB}
\Sigma^B_p(\lambda) = (ig)^2 \sum\limits_{|k|>1,|q|<1} G_{p-k+q}(\lambda+iE_{kq})
\ee
\begin{equation}\label{SlambdaB}
\sigma^B_p(\lambda_1,\lambda_2) = g^2\sum\limits_{|s|<1,|k|>1} W_{p-s+k}(\lambda_1+iE_{sk},\lambda_2-iE_{sk})
\end{equation}

At this point it is worthwhile emphasizing the consistency of our approach regarding the sum rule,
\be\label{balance}
\sum_p n_p(\tau) = 1\,,
\ee
which follows form definition \eqref{prob}.
If one used inconsistent approximations for $\Sigma_p$ and $\sigma_p$ it could happen that this condition would be violated in some orders in $g$. We stress that if $\Sigma_p$ and $\sigma_p$ are determined self-consistently from a single expression $\hat{\Sigma}$ this will not happen. Indeed, from Eq. \eqref{W} we can obtain the following kinetic-like equation:
\be\label{balans}
\frac{dW_p(\tau,\tau)}{d\tau} = \int\limits_0^{\tau}dt \Sigma_p(\tau-t)W_{p}(t,\tau) +{\rm h.c.}\\
\displaystyle
+ \int\limits_0^{\tau}dt  \sigma_p(t,\tau)G_p(\tau-t)
+{\rm h.c.}
\ee
It is more insightful to consider this equation in dual space. Namely, let us pick some quantity $X_p$ and consider evolution of its average $\langle X \rangle \equiv \sum_p X_p n_p(\tau)$. Using this equation with self-energy parts given by \eqref{22} and \eqref{23} we obtain:
\be
\frac{d \langle X \rangle}{d\tau}= \sum\limits_{N=1}^{\infty}(ig)^{N+1}S_N(X)+ \sum\limits_{N=1}^{\infty}(-ig)^{N+1}S^*_N(X)\,
\ee
with
\begin{multline}
\hspace{-5mm}S_N(X) =\sum\limits_{p,|k|>1,|q_i|<1}(X_p-X_{p-k+q_1}) \int\limits_{\Delta_{\tau}}\prod_{i=1}^{N+1}dt_i  \,e^{-iE_{kq_1}t_1}\times \\ \times G_{p-k+q_1}(t_1)\dots
G_{p-k+q_N}(t_N)e^{-iE_{kq_N}t_N} W_{p}(t_{N+1},\tau)\,.
\end{multline}
This, in particular, shows that balance, \eqref{balance}, is conserved for any moment in time (to see this one should put $X_p=1$). The balance property, \eqref{balance}, holds for any choice of $\hat{\Sigma}$.

In the next section we
derive approximate expressions for the self-energy contributions and solve \eqref{Glambda} and \eqref{Slambda} in a way to ensure that the sum rule, \eqref{balance}, is satisfied up to order $O(g^2)$.

\section{\label{Solution}Asymptotic distribution}

\subsection{Feynman Green function}

To find the probability distribution function we first have to solve Eq. \eqref{Glambda} to determine Feynman Green function. To do this we first
take the following expression in the leading order in $g$:
\be\label{Glambda2}
G_p(\lambda)^{-1} = \lambda + g^2 \sum\limits_{|k|>1,|q|<1} \frac{1}{\lambda+iE_{kq}}\,.
\ee
This function has a cut the complex plane $\lambda$ along imaginary axis. We can introduce spectral function $A_p$ as:
\be
G_p(\lambda) = \int dz\frac{A_p(z)}{\lambda + iz}\,,
\ee
then time dependence will be given as:
\be
G_r(t) = \int d\omega A_{r}(\omega) e^{-i\omega t}\,.
\ee
From formula \eqref{Glambda2} one can easily conclude that:
\be
A_p(\omega) = \frac{1}{\pi} \frac{\Gamma_p(\omega)/2}{(\omega-S_p(\omega))^2 + (\Gamma_p(\omega)/2)^2}\,,
\ee
where
\be\label{GG}
\Gamma_p(\omega) = 2\pi \gamma^2 \left(\frac{2\pi}{L}\right)^2\sum_{|k|>1,|q|<1}\delta(\omega-E_{kq})\,,
\ee
and
\be
S_p(\omega) = {\rm v.p.} \int \frac{dE}{2\pi} \frac{\Gamma_{p}(E)}{\omega-E}\,,
\ee
where $v.p.$ stands for principal value. 
Remind also that $\gamma = gL/(2\pi)$. Function $\Gamma_p(\omega)$ is zero below some threshold.
For $|p|>q_0\equiv {\rm min}(1,\eta)$ this threshold is negative so $A_p(\omega)$ has the shape of a Lorentz distribution centered approximately at $\omega=0$, with width $\Gamma_p \equiv \Gamma_p(0)$. This form of the spectral function is typical for decay processes, which, in our case, reflect the possibility of the impurity scattering on the host particle and losing momentum.
The inverse width determines the time scale prior to which $G_p(t)$ has diffusive dynamics. Namely, for:
\be\label{cT}
1\ll t \lesssim \frac{1}{\gamma^2}\log\frac{1}{\gamma^2}\,,
\ee
\be\label{G1}
G_p(t) = e^{-\frac{\Gamma_p}{2} t}\,.
\ee
On the other hand, for $|p|<q_0$ the main contribution comes from the domain where $\Gamma_p=0$; in this case we can replace the spectral function with
$A_p(\omega) = \delta(\omega-S_p(\omega))$, which gives some oscillatory contribution that is not important for our consideration for times satisfying \eqref{cT}; therefore, for $|p|<q_0$ we have
\be
G_p(t) = 1\,.
\ee
Even though these naive dynamics might acquire some subdiffusive corrections \cite{Zvonarev2007},\cite{PhysRevA.80.011603},\cite{Khodas2007fermi} they are important beyond the time scale, \eqref{cT}. Here we would like to stress that we consider not genuine $t\to\infty$, but a large enough time, meaning that $e^{-\gamma^2t}\sim \gamma^2$ which seems to be appropriate in the $\gamma\to 0$ case.

The width $\Gamma_p$ can be easily calculated from its definition, \eqref{GG}, which is nothing but the Fermi Golden Rule. So, for $p>q_0$ we have
\be\label{Fermi}
\frac{\Gamma_p}{2\pi \gamma^2} =
\left\{\begin{array}{ll}
\theta(1-p)\log\frac{1+\eta}{1-\eta}+\theta(p-1)\log\frac{p+\eta}{p-\eta}, & \eta<1 \\
\log\left|\frac{p+\eta}{p-\eta}\right|-\theta(\eta-p)\log\frac{\eta+1}{\eta-1},&\eta>1 \\
\log\frac{p+1}{p-1},&\eta=1
\end{array}\right.
\ee
The time domain, \eqref{cT}, in $\lambda$ space can be easily expressed as
\be
\frac{\gamma^2}{\log\gamma^{-2}}\lesssim |\lambda| \ll 1\,,
\ee
which, in practice, means that for $\Sigma_p(\lambda)$ in \eqref{Glambda} one can consider $\lambda \to 0$ keeping it only to regularize possible divergences in the denominator. Performing calculations analogous to the Boltzmann case we see that ladder contributions do not change result \eqref{G1} in the leading order.
Moreover, for equal mass one can confirm the result for the spectral function directly from the Bethe ansatz \cite{Burovski2014}.
The only places where higher orders in $\gamma$ may play a role are vicinities of the discontinuities of the Fermi Golden Rule result, \eqref{Fermi}, $p=1$, $p=\eta$, and probably some other countable set of points (see \cite{BChGL2013,GLCh2014}). But this consideration is beyond the scope of this paper.

\subsection{Generalized distribution function (Boltzmann case)}

Once we have determined the Green function as
\be\label{GGG}
G_p(\lambda) = \frac{1}{\lambda+ \Gamma_p/2}\,,
\ee
we can solve the QBE, \eqref{QBE}. For the bubble diagrams (Fig. \ref{LadderB}) QBE reads
\begin{multline}\label{BW}
W_{p}(\lambda_1,\lambda_2) = \frac{n_p^{(0)}}{(\lambda_1+\Gamma_p/2)(\lambda_2+\Gamma_p/2)}+\\
+g^2\sum\limits_{|s|<1,|k|>1} \frac{W_{p-s+k}(\lambda_1+iE_{sk},\lambda_2-iE_{sk})}{(\lambda_1+\Gamma_p/2)(\lambda_2+\Gamma_p/2)}\,.
\end{multline}
The asymptotic distribution is given by the formula
\be
n_p^{\infty}=W_p(\lambda_1,\lambda_2)^{\rm L.T.}\,,
\ee
where ${\rm L.T.}$ stands for Laplace transformation, namely,
\be
\label{inf}
W_p(\lambda_1,\lambda_2)^{\rm L.T.} \equiv \lim\limits_{t\to\infty}\int_C \frac{d\lambda_1}{2\pi i } \int_C \frac{d\lambda_2}{2\pi i}e^{t(\lambda_1+\lambda_2)} W_p(\lambda_1,\lambda_2)\,.
\ee
Here $C$ is the contour that goes from $-i\infty$ to $i\infty$ and lies to the right of all the singularities of the integrand, which, in our case, means ${\rm Im}\lambda > 0$; $t\to\infty$ represents the right edge of the time domain, \eqref{cT}. The initial distribution, which, we, for simplicity, assume to be $n_p^{(0)}=\delta_{p,p_0}$, can be easily extended to an arbitrary diagonal due to linearity.

The structure of the QBE immediately suggests that the solution can be obtained by an iterative procedure. Because balance is conserved \eqref{balance}, it is reasonable to make iterations till
\be\label{bb}
\sum_pn^{\infty}_p=1
\ee
in the limit $\gamma\to 0$. For example, if the initial momentum $|p_0|<q_0$, then even a "zero" iteration term will do the job. Namely,
\be
W^{(0)}_p(\lambda_1,\lambda_2) = \frac{1}{\lambda_1+\Gamma_p/2}n_p^{(0)}\frac{1}{\lambda_2+\Gamma_p/2}\,
\ee
gives
\be
n^{\infty}_{p} = n_p^{(0)} \lim\limits_{t\to\infty} e^{-t\Gamma_p}\,.
\ee
But for $p=p_0<q_0$ we have $\Gamma_p=0$ and condition \eqref{bb} is saturated, providing $n^{\infty}_{p} = n_p^{(0)}$. This merely shows
the kinematic impossibility of single scattering due to Pauli blocking \cite{BChGL2013,Lychkovskiy2013}.

Assume now that the initial momentum $p_0>q_0$ then $\Gamma_{p_0}>0$ and the contribution of the "zero" iteration term is negligible, while the first iteration gives
\begin{widetext}
\be
\label{f1}
 W^{(1)}_p(\lambda_1,\lambda_2) =   \frac{1}{\lambda_1+\Gamma_p/2}\left[g^2\sum\limits_q\sum\limits_{|s|<1,|k|>1}\frac{n^{(0)}_q\delta_{q,p-s+k}\theta(|q-p+s|-1)}{(\lambda_1+iE_{sk}+\Gamma_q/2)(\lambda_2-iE_{sk}+\Gamma_q/2)}\right]\frac{1}{\lambda_2+\Gamma_p/2}\,.
\ee
\end{widetext}
One can easily understand that each iteration will give just the sum of the inverse polynomials in $\lambda_i$. The long-time limit \eqref{inf} means that only residues at $\lambda_i=0$ are important. One can also show that these poles are simple, so we can ignore all other $\lambda_i$ dependence by evaluating the corresponding functions at $\lambda_i=0$. Therefore, since $\Gamma_q=\Gamma_{p_0}>0$ we can safely put $\lambda_1=\lambda_2=0$ in the square brackets in \eqref{f1}. After that we perform summation over $s$ and $k$.  The only way to get a non vanishing result at $\gamma\to 0$ is to make the zeroes of $E_{sk}$ lie in the summation domain. This will lead to the restriction that momenta $q$ and $p$ must lie in a certain domain, $(q,p)\in\Omega$, which we specify below (cf. \cite{BChGL2013,GLCh2014}).
This way, we get
\be\label{f1.1}
 W^{(1)}_p(\lambda_1,\lambda_2) =   \frac{1}{\lambda_1+\Gamma_p/2}\left(\sum_qn^{(0)}_q\mathcal{P}^{(1)}_{q\to p}\right)\frac{1}{\lambda_2+\Gamma_p/2}\,,
\ee
with
\be\label{PPP}
\mathcal{P}^{(1)}_{q\to p} = \frac{2\pi}{L}2\pi\gamma^2\frac{\theta_{\Omega}(q,p)}{|p-q|\Gamma_{q}}\,.
\ee
Here the step function $\theta_{\Omega}(q,p)$ means that point $(q,p)$ lies in the domain $\Omega$. More specifically, this means that:
\begin{multline}\label{Domain}
\theta_{\Omega}(q,p)\equiv\\\theta\left(\left|q\frac{1+\eta}{2\eta}+p\frac{1-\eta}{2\eta}\right|-1\right)\theta\left(1-\left|q\frac{1-\eta}{2\eta}+p\frac{1+\eta}{2\eta}\right|\right)\,.
\end{multline}
If the initial momentum is such that all $|p|$ satisfying condition \eqref{Domain} are less than $q_0$, then this iteration gives final answer:
\be\label{first}
n^{\infty}_p=[n^{\infty}_p]^{(1)} = \theta(q_0-|p|)\sum_qn^{(0)}_q\mathcal{P}^{(1)}_{q\to p}
\ee
This happens for $|p_0|<q_1$, with $q_1=\frac{3\eta-1}{\eta+1}$ for $\eta>1$ and $q_1=\frac{\eta(3-\eta)}{\eta+1}$ for $\eta<1$.
Kinematically $[q_0,q_1]$ is the range of momenta within which the impurity momentum drops below $q_0$ in one scattering \cite{GLCh2014}.
In the general case we must reiterate till $|p|<q_0$. If this happens after $n$ iteration, then the corresponding probability distribution reads
\be
n^{\infty}_p = \theta(q_0-|k|) \sum\limits_{j=1}^n \mathcal{P}^{(j)}_{p_0\to k}
\ee
where
\be
{\cal P}^{(j)}_{ p_0\rightarrow k}=\sum_{|q|>q_0}{\cal P}^{(1)}_{ p_0\rightarrow q} {\cal P}^{(j-1)}_{ q\rightarrow k}.
\ee
If the initial momentum is below $q_{\infty}={\rm max}(1,\eta)$, then we need finite number of iterations; if not, the exact answer is given only by an infinite number of iterations, though the approximation error is very well controlled \cite{GLCh2014}.

In the case of equal masses it turns out that the first iteration gives full (valid for any initial momentum) asymptotic solution of Eq. \eqref{first}, which reads
\be\label{B1}
n^{\infty}_p = \frac{2\pi}{L}\frac{1}{\log\frac{p_0+1}{p_0-1}}\frac{\theta(1-|p|)}{p_0-p} ,\,\,\,\, p_0>1.
\ee
We would like to emphasize that this result was obtained in the lowest possible approximation for the self-energy (Fig. \ref{LadderB}); in the next subsection
we show that ladder diagrams modify this result substantially even in the leading $\gamma$ order.

Also, let us comment on how to derive the usual Boltzmann equation to describe not only asymptotic distributions but
time dependence upon the approach of this asymptotic as well. Note that in the time domain Eq. \eqref{BW} takes the following form
\begin{multline}\label{WNL}
W_{p}(t_1,t_2) =  \left(n_p^{(0)}+g^2\sum\limits_{|s|<1,|k|>1}\int\limits_0^{t_1}d\tau_1\int\limits_0^{t_2}d\tau_2e^{\frac{\Gamma_p\tau_1}{2}}\times\right. \\
  \left. \times e^{-iE_{sk}(\tau_1-\tau_2)}W_{p-s+k}(\tau_1,\tau_2)e^{\frac{\Gamma_p\tau_2}{2}}\right)e^{-\frac{(t_1+t_2)\Gamma_p}{2}}
\end{multline}
From this equation we get
\begin{multline}\label{ww}
\frac{dn_p(t)}{dt} \equiv \frac{dW_p(t,t)}{dt} = -\Gamma_p n_p(t) + \\+g^2 \sum\limits_{|s|<1,|k|>1} \int\limits_{0}^{t}d\tau e^{-iE_{sk}\tau-\gamma^2\Gamma_p\tau} W_{p-s+k}(t-\tau,t) \\ \displaystyle+ g^2 \sum\limits_{|s|<1,|k|>1} \int\limits_{0}^{t}d\tau e^{iE_{sk}\tau-\gamma^2\Gamma_p\tau} W_{p-s+k}(t,t-\tau).
\end{multline}
Now assuming weak dependence of the $W_p(t_1,t_2)$ on the difference between $t_1$ and $t_2$ we may put $W_{p}(t,t-\tau)\approx W_{p}(t-\tau,t)\approx n_p(t)$, therefore equation \eqref{ww} at $t\to\infty$ will take following form:
\be
\frac{dn_p(t)}{dt} = -\Gamma_p n_p(t) + g^2 \sum\limits_{|s|<1,|k|>1} 2\pi \delta(E_{sk}) n_{p-s+k}(t).
\ee
Furthermore, using notations \eqref{PPP},\eqref{Domain}  we can present it as:
\be\label{FermiF}
\frac{dn_p(t)}{dt} = -\Gamma_p n_p(t) + \sum\limits_{q} \Gamma_{q\to p} n_{q}(t),
\ee
where
\be\label{Fe}
\Gamma_{q\to p}  =  \frac{2\pi}{L}2\pi \gamma^2\frac{ \theta_{\Omega}(q,p)}{|p-q|},
\ee
Obviously $\sum_q \Gamma_{p\to q} = \Gamma_p$.
The Boltzmann equation in the form of \eqref{FermiF} was the starting point in Ref. \cite{GLCh2014}, where it was derived directly from the Fermi Golden Rule treatment of the quantum mechanical transition probability.

\subsection{Generalized distribution function (Ladder case)}

In the previous subsection we considered an asymptotical solution based on the self-energy given by a bubble diagram (Fig. \ref{LadderB}). The ladder effects change the QBE by introducing denominators as in \eqref{Slambda}.
To estimate those effects we can check how the residue at the pole $\lambda_i=0$ is affected by the presence of corresponding denominators.
Then the first iteration result of Eq. \eqref{f1} is modified by multiplication of the summand on the $|f_{p,s}|^2$, where:
\be\label{f2}
 f_{p,s} = \left(1-g \sum\limits_{|q|<1} \frac{1}{E_{sq}-i0}\right)^{-1}.
\ee
Since $|s|<1$ the expression in the denominator will always have an imaginary part:
\begin{multline}
 f_{p,s} = \left(1+(\dots)+i\pi g \sum\limits_{|q|<1}\delta(E_{sq})\right)^{-1} = \\=\left(1+(\dots)+i2\pi \eta \frac{\gamma}{|\eta-1|}\frac{1+\theta(1-|s-2p|)}{|p_0-p|} \right)^{-1}\,.
\end{multline}
Here the dots denote some irrelevant real part. We see that the Boltzmann description, \eqref{BW}, is valid and all results from the previous subsection are correct if
\be
\frac{\gamma}{|\eta-1|} \ll 1\,.
\ee
On the other hand, in the equal masses case ($\eta=1$) the contribution of the residues at poles of Green functions is negligible. This way, for $\eta=1$ probability distribution function is completely determined by the residues at poles of $\sigma_p^L(\lambda_1,\lambda_2)$ \eqref{Slambda} that come from ladder contributions. They are responsible for multiple scattering events.

\begin{figure*}[t]
\includegraphics[width= 0.8\linewidth]{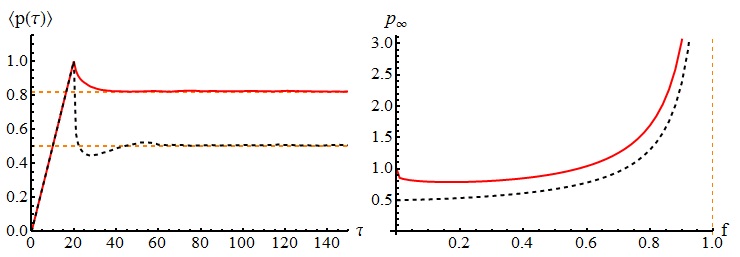}
\caption{\label{dynamic}
Left: Average momentum of the impurity vs time in the equal-mass case for the applied external force $f=0.05$.
Right: Steady-state momentum vs applied force obtained by formulas \eqref{pL} (solid red line) and \eqref{pB} (dashed black line). The steady-state momentum diverges at a critical force $f_{c1}=1$ (represented by the dashed vertical line).
In both panels to solid red line corresponds to the correct (ladder) kinetic equation, while black the dashed line shows the results of the Fermi Golden Rule kinetic equation.
}
\end{figure*}

Henceforth we focus on the equal-mass case, $\eta=1$. Analogously to the previous subsection let us consider the initial momentum $p_0>q_0=1$. After the first iteration, instead of \eqref{f1} one can get
\begin{widetext}
\be\label{wWw}
W^{(1)}_{p}(\lambda_1,\lambda_2) = g^2\sum\limits_{|s|<1,|k|>1} \frac{G_p(\lambda_1)G_{p-s+k}(\lambda_1+iE_{sk})G_{p-s+k}(\lambda_2-iE_{sk})G^*_p(\lambda_2)\delta_{p-s+k,p_0}}{\left(1-ig \sum\limits_{|q|<1} G_{p-s+q}(\lambda_1+iE_{sq})\right)\left(1+ig \sum\limits_{|q|<1} G^*_{p-s+q}(\lambda_2-iE_{sq})\right)}\,.
\ee
\end{widetext}
Here, for the sake of brevity, we use notation $G_p(\lambda)$ defined in Eq. \eqref{GGG}.
Now taking into account that the final distribution is determined after Laplace transformation at equal times, \eqref{inf}, we may shift $\lambda_1\to\lambda_1+i\xi$ and
$\lambda_1\to\lambda_1-i\xi$ for any real $\xi$. Moreover, we find it simpler not to consider the probability distribution function but use dual description of some quantity $X_p$ and its average:
\be\label{average0}
\langle X \rangle = \sum_p X_p n_p^{\infty}
\ee
Using formula \eqref{wWw} we obtain:
\begin{widetext}
\be\label{average1}
\langle X \rangle =g^2\sum\limits_{|s|<1,|k|>1}\left[ \frac{X_{p_0+s-k}G_{p_0+s-k}(\lambda_1+i(k-p_0)(k-s))G_{p_0}(\lambda_1)G_{p_0}(\lambda_2)G^*_{p_0+s-k}(\lambda_2-i(k-p_0)(k-s))}{\left(1-ig \sum\limits_{|q|<1} G_{p_0-k+q}(\lambda_1+i(k-p_0)(k-q))\right)\left(1+ig \sum\limits_{|q|<1} G^*_{p_0-k+q}(\lambda_2-i(k-p_0)(k-q))\right)}\right]^{\rm L.T.}\,.
\ee
\end{widetext}
Remember that ${\rm L.T.}$ means that in order to obtain $\langle X\rangle$ one has to perform Laplace transformation on both $\lambda_1$ and $\lambda_2$ at the same time and then send this time to infinity, similar to Eq. \eqref{inf}.  The dominator in the sum in \eqref{average1} vanishes at $\lambda$ equals to
\be
i\lambda = (k-p_0) \left(k-\coth \left(\frac{k-p_0}{2 \gamma}\right)\right)
\ee
Calculating residues at this point we obtain from Eq. \eqref{average1} the following answer:
\begin{multline}\label{average2}
\hspace{0pt minus 12.2pt}\langle X \rangle =\left(\frac{2\pi}{L}\right)^2\mathop{\sum\limits_{|s|<1}}_{|k|>1} \frac{X_{p_0+s-k}}{\left(s-\coth\frac{k-p_0}{2\gamma}\right)^2\left(1-\cosh\frac{k-p_0}{\gamma}\right)^2} \times\\ \times \frac{(k-p_0)^2}{(k-p_0)^2\left(k-\coth\frac{k-p_0}{2\gamma}\right)^2+(\Gamma_{p_0}/2)^2}
\end{multline}
The integral over $k$ acquires its value from the small domain around $k_*$, which is the solution of the equation:
\be
k_*=\coth\frac{k_*-p_0}{2\gamma}\,,
\ee
which, at $\gamma\to0$, can be written as
\be
k_* = p_0+2\gamma \coth^{-1} p_0 = p_0+\frac{\Gamma_{p_0}}{2\pi \gamma} \,.
\ee
Now expanding the integrand near this value and assuming that $X_s$ is a smooth function of momentum we get:
\begin{multline}
\langle X \rangle \overset{\gamma\to0}{=}\left(\frac{2\pi}{L}\right)^2\sum\limits_{|s|<1,|k|>1}\frac{X_s}{(p_0-s)^2}\frac{(p_0^2-1)^2}{4} \times \\ \times
 \frac{\gamma^2}{(k-k_*)^2(p_0^2-1)^2/4+(\pi \gamma^2)^2} = \\ =\frac{2\pi}{L}\frac{p_0^2-1}{2}\sum\limits_{|s|<1} \frac{X_s}{(p_0-s)^2}.
\end{multline}
So comparing with formula \eqref{average0} we obtain
\be
n_p^{\infty} =  \frac{2\pi}{L}\frac{p_0^2-1}{2} \frac{\theta(1-|p|)}{(p_0-p)^2}
\ee
One can note how different this answer is from result \eqref{B1}, which disregards ladder effects. Therefore,
we see that multiple scattering events of the impurity on host particles change the probability distribution function even
in the leading order. In contrast, the Feynman Green function remains practically the same.

Performing calculations similar to those in the previous subsection we can obtain analog of the Boltzmann equation \eqref{FermiF}:
\be\label{BE}
\frac{dn_p(t)}{dt} = -\Gamma_p n_p(t) +\sum\limits_{q} \tilde{\Gamma}_{q\to p} n_{q}(t),
\ee
where
\be
\tilde{\Gamma}_{q\to p}  =  \frac{2\pi}{L}\Gamma_q\frac{q^2-1}{2} \frac{\theta(1-|p|)}{(q-p)^2},
\ee
Obviously, $\sum_q \tilde{\Gamma}_{p\to q} = \Gamma_p$. So this equation looks like ordinary Boltzmann equation; the only difference is
that transition rates $\tilde{\Gamma}_{p\to q}$ are no longer determined from the Fermi Golden Rule \eqref{Fe}, even though the total width remains unchanged.

\section{\label{Force}Force}

In Ref. \cite{GLCh2014} the authors considered how a constant force applied to an impurity affects the dynamics in the case of light, $\eta<1$, and heavy,  $\eta>1$, impurities.
Now using the correct Boltzmann equation, \eqref{BE}, we can also investigate the $\eta=1$ case. One can generalize Eq. \eqref{BE}
to account for the constant force $F$ acting on the impurity:
\be\label{BE1}
\frac{\partial n_k(t)}{\partial t} + F\frac{\partial n_k(t)}{\partial k} = -\Gamma_k n_k(t) + \sum\limits_{q} \tilde{\Gamma}_{q\to k} n_{q}(t)
\ee
Without loss of generality we assume $f>0$. In this case Eq. \eqref{BE1} allows us to put $n_k=0$ for $k<-1$. For the other values of $k$ we can introduce following notations
\be
n_k = \frac{2\pi}{L}\left\{\begin{array}{ll}
\nu(k), & k\in[-1,1] \\
\chi(k),  & k>1
\end{array}\right.\,,
\ee
where we have performed certain rescalings. It is also convenient to rescale time and force:
\be
f= \frac{F}{2\pi\gamma^2},\,\,\, \tau = 2\pi \gamma^2 t\,.
\ee
Then Eq. \eqref{BE1} reads
\begin{multline}\label{chi}
\frac{\partial \chi(k,\tau)}{\partial \tau} + f\frac{\partial \chi(k,\tau)}{\partial k} = - \log\frac{k+1}{k-1}\chi(k,t)\,,\\
\frac{\partial \nu(k,\tau)}{\partial \tau} + f\frac{\partial \nu(k,\tau)}{\partial k} = \int\limits_1^{\infty} dq \chi(q,\tau)\frac{q^2-1}{2(q-k)^2}\log  \frac{q+1}{q-1}\,.
\end{multline}
The dynamics of the average momentum defined as
\be\label{pppP}
\langle p(\tau) \rangle = \int\limits_{-1}^{1} dk \, k \nu(k,\tau) + \int\limits^{\infty}_{1} dk \, k \chi(k,\tau)
\ee
is shown in the left panel in Fig. \ref{dynamic}. It is compared to the dynamics that comes from the Boltzmann equation with transition rates determined by the Fermi Golden Rule, \eqref{FermiF}. We see that in both cases the system quickly reaches the steady state. However, the momenta obtained from the correct kinetic equation, which takes into account ladder effects, are larger.
The steady state is characterized by the distribution function
\be\label{chi2}
\chi(k) = \exp\left(-\frac{1}{f}\left((k-1)\log\frac{k+1}{k-1}+2\log\frac{k+1}{2})\right)\right)\,.
\ee
up to some normalization constant. This function gets finite normalization for $f<f_{c0}=2$. This way, $f_{c0}$ is a critical force for which the steady state exists. The corresponding average momentum reads
\begin{multline}\label{pL}
p_{\infty}^L = \frac{\int\limits_1^{\infty}dq \chi(q) \left(\frac{3q^2-1}{2}\log\frac{q+1}{q-1} -2 q\right)}{\int\limits_1^{\infty}dq \chi(q) \left(q\log\frac{q+1}{q-1} - 1\right)}\\=\frac{1-\frac{2-3f}{f}\int\limits_1^{\infty}dq \, q \chi(q)}{1-\frac{1-f}{f}\int\limits_1^{\infty}dq \, \chi(q)}.
\end{multline}
Note that if one were to consider the bubble diagram only, one would get the following answer:
\be\label{pB}
p_{\infty}^B = \frac{\int\limits_1^{\infty}dq \, q \chi(q)}{2\int\limits_1^{\infty}dq \, \chi(q)}\,.
\ee
All these expressions are finite for $f<f_{c1}=1$.
These two results are compared at Fig. \ref{dynamic}. We see that calculations that take into account ladder contribution give higher steady state momentum.
Unfortunately at small forces $f\ll \gamma^2$ our results are still hardly applicable because the average momentum is equal to the Fermi momentum where
our simple approximation for the Feynman Green function, \eqref{GGG}, is not valid, and more careful solution of the Dyson equation needed.
We hope to clarify this issue in future.

\section{\label{Discussion} Summary and discussion}

To summarize, we have systematically derived the kinetic equation of an impurity in a Tonks-Girardeau gas in the weak-coupling regime.
We have rederived some of the results in \cite{GLCh2014} when the masses of the host particle and impurity are different, confirmed the applicability of the approach used there, and generalized the description to the equal-mass case. Our QBE correctly takes into account multiple coherent scatterings and
reproduces results that follows from the Bethe ansatz treatment \cite{BChGL2013,Burovski2014}. At equal masses we have also derived a Boltzmann like kinetic equation
which is absolutely new. This has allowed us to consider a constant external force applied to the impurity and calculate the steady-state
momentum. We have shown that application of the naive Boltzmann equation with the Fermi Golden Rule transition rates gives a completely wrong answer in the
equal-mass case.

The perturbative description of the system developed here is a powerful and versatile approach.
It allows straightforward generalization for the arbitrary weak interaction between host and particle and for quite a general host, which is
the subject of further investigations. Another important possible generalization is to consider the system at finite temperature with at arbitrary trap potential. We believe that our approach will be fruitful in these cases as well. From the theoretical point of view it is also interesting to consider
next-to-leading-order corrections. This looks the most challenging issue, because one not only must find the exact solution of the QBE but also must
take into account other diagrams not considered here. Nevertheless, the approach described in this manuscript allows to do this systematically
which we believe will be done in the very near future.

\acknowledgements{The author is grateful to V. Cheianov, O. Lychkovskiy,
E. Burovskiy, M. Zvonarev and L. Glazman  for fruitful discussions.
The present work was supported by ERC Grant No. 279738-NEDFOQ.}

\bibliography{Keldysh}

\end{document}